\documentclass{aastex631}

\usepackage{multirow}

\begin{document}

\title{Energy-Dependent Analyses of the Gamma-Ray Emission from HESS J1857+026 with {\it Fermi}-LAT}

\author[0009-0004-5450-8237]{Xiaolei Guo}
\author{Xi Liu}
\author[0000-0001-5135-5942]{Yuliang Xin}
\affiliation{School of Physical Science and Technology, Southwest Jiaotong University, Chengdu 610031, People’s Republic of China; \href{mailto:xlguo@swjtu.edu.cn}{xlguo@swjtu.edu.cn}}



\begin{abstract}
We report the discovery of energy-dependent morphology for the GeV gamma-ray emission from HESS J1857+026 with more than 13 years of {\it Fermi} Large Area Telescope (LAT) data.
The GeV gamma-ray emission from this region is composed of two extended components. The hard component with an index of $1.74 \pm 0.07$ in the energy range of 0.5-500 GeV is spatially coincident with HESS J1857+026, and its 68\% containment radius varies from $\sim 0.44^\circ$ below 40 GeV to $\sim 0.30^\circ$ above 140 GeV. 
The hard GeV gamma-ray spectrum and the energy-dependent morphology of HESS J1857+026 make it favor a PWN origin, which is associated with the energetic pulsar, PSR J1856+0245.
The soft component with an index of $2.70 \pm 0.16$ and another extended gamma-ray source detected in this region, 4FGL J1857.9+0313e with an index of $2.55 \pm 0.07$, are spatially coincidence with two molecular clumps in the northeast and southwest of HESS J1857+026, which favors the hadronic process, and the protons could be accelerated by the hypothetical SNR associated with PSR J1856+0245.

\end{abstract}



\section{Introduction} \label{sec:intro}

More than two hundred of very-high-energy (VHE; $>$ 100 GeV) gamma-ray sources have been detected with the operations of ground-based Cherenkov telescopes,
such as H.E.S.S. \citep{2004APh....22..109A}, MAGIC \citep{2016APh....72...61A}, HAWC \citep{2013APh....50...26A} and LHAASO \citep{2023arXiv230517030C}. 
The gamma-ray emission could be produced by the hadronic interactions in which $\pi^0$ decays into two gamma-ray photons subsequently, via the inverse Compton (IC) scattering process or through non-thermal bremsstrahlung radiation from high energy electrons.
Most of the VHE gamma-ray sources have been identified to be pulsar wind nebulae (PWNe), supernova remnants (SNRs), X-ray binaries, TeV halos and so on. 
However, some TeV gamma-ray sources are still unidentified, and the multi-wavelength studies of these sources are crucial for revealing their nature and probing the origin of 
cosmic rays. 

HESS J1857+026 was first discovered as a VHE gamma-ray source 
with an extension of $(0.11 \pm 0.08)^\circ \times (0.08 \pm 0.03)^\circ$ \citep{2008A&A...477..353A}.
In \citet{2018A&A...612A...1H}, the morphology of HESS J1857+026 was described by a two dimensional (2D) Gaussian component with an approximate size of 
$0.26^\circ \pm 0.06^\circ$.
\citet{2008ApJ...682L..41H} discovered an energetic pulsar, PSR J1856+0245, in the direction of HESS J1857+026, which makes HESS J1857+026 to be a potential PWN candidate.
The period and spin-down luminosity of PSR J1856+0245 are P = 81 ms and $\dot E$ = $4.6 \times 10^{36}$ erg s$^{-1}$, respectively, with the characteristic age of $\tau_c$ = 21 kyr.
The distance of PSR J1856+0245 was first estimated to be $\sim$ 9 kpc, derived by the dispersion measure (DM) with electron density model of \cite{2002astro.ph..7156C}. 
And, an updated distance of $\sim 6.3$ kpc is given by ATNF Pulsar Catalog \footnote{https://www.atnf.csiro.au/research/pulsar/psrcat/}
according to the electron density model of \cite{2017ApJ...835...29Y}.

MAGIC carried out follow-up observations and presented energy-dependent morphology of this region \citep{2014A&A...571A..96M}.
In the energy range of 300 GeV-1 TeV, the morphology observed by MAGIC is compatible with H.E.S.S. observation.
However, two separate gamma-ray sources, named MAGIC J1857.2+0263 (hereafter MAG1) and MAGIC J1857.6+0297 (hereafter MAG2), 
were detected with the data above 1 TeV.
MAG1 is an extended source with the intrinsic extension of $(0.17 \pm 0.03_{\rm stat} \pm 0.02_{\rm sys})^\circ \times (0.06 \pm 0.03_{\rm stat} \pm 0.02_{\rm sys})^\circ$, while MAG2 is compatible with a point source.
\citet{2014A&A...571A..96M} interpreted MAG1 as a PWN powered by PSR J1856+0245, and MAG2 may be associated with a molecular cloud complex containing an H{\sc ii} region located at $\sim$ 3.7 kpc and a possible gas cavity. 
Another pulsar, PSR J1857+0300, was discovered in the direction of MAG2 \citep{2017ApJ...834..137L}. 
The characteristic age and spin-down luminosity of PSR J1857+0300 are $\tau_{\rm c}$ $\sim 4.6\times 10^{6}$ yr and $\dot E$ $\sim 2.3 \times 10^{32} \ {\rm erg}\ {\rm s^{-1}}$, with the distance of $\sim 6.7$ kpc.
Meanwhile, an elliptical superbubble, was detected with neutral gas observation \citep{2021A&A...652A.142P}, 
which is also spatially coincident with HESS J1857+026. The kinematical distance of the superbubble is about 5.5 kpc and 
is close to the DM distance of PSR J1856+0245 ($\sim 6.3$ kpc). 
\cite{2021A&A...652A.142P} concluded that the TeV emission of HESS J1857+026
originates from the superbubble, and PSR J1856+0245 is located inside the superbubble.
In addition, they found five molecular components in the velocity interval between 78 and 90 $\rm{km}\ \rm{s}^{-1}$
with $^{13}\rm{CO}$(J=1-0) observations, which are probably associated with the superbubble.
And they favor a single gamma-ray source scenario instead of the superposition of two gamma-ray 
sources. \cite{2021A&A...652A.142P} also performed radio observations at 1.5 GHz and 6.0 GHz with VLA. Nevertheless, 
no significant radio emission was detected in this region.

In the energy range of 1-25 TeV, the Water Cherenkov Detector Array (WCDA) of Large High Altitude Air Shower Observatory (LHAASO) detected an extended gamma-ray source, 1LHAASO J1857+0245, which is spatially consistent with HESS J1857+026 \citep{2023arXiv230517030C}.
The spatial template of 1LHAASO J1857+0245 is described by a 2D Gaussian with $\sigma = 0.24^\circ \pm 0.04^\circ$, and its gamma-ray spectrum in 1-25 TeV is modeled by a power law with an index of 2.93 $\pm$ 0.07.

The GeV gamma-ray emission from HESS J1857+026 was first detected using a point source hypothesis by {\it Fermi}-LAT \citep{2009ApJ...697.1071A}, 
while no gamma-ray pulsation from PSR J1856+0245 was observed \citep{2012A&A...544A...3R}. 
Since no obvious X-ray emission was detected, only an upper limit of $F_{1-10\ {\rm keV}} = 2 \times 10^{-12}$ erg cm$^{-2}$ s$^{-1}$
was obtained \citep{2012A&A...544A...3R}. 
Considering the uncertain origin and the complexity of this region, the detailed analysis with more GeV observational data will be helpful to investigate 
the origin of the gamma-ray emission.

Taking advantage of more than 13 years of {\it Fermi}-LAT data, we performed an energy-dependent analyses of the region around HESS J1857+026 
and discussed the nature of this source.
The paper is organized as follows. In Section 2, we describe the data analysis routines and present our results. 
In Section 3, we discuss the radiation mechanisms and the nature of HESS J1857+026. 
And the conclusion for this work is presented in Section 4.

\section{Fermi-LAT Data and Results}\label{sec:fermi-data}

The {\it Fermi}-LAT Pass 8 data we analyzed are collected from August 4, 2008 to March 4, 2022 with energies from 500 MeV to 500 GeV.
The region of interest (ROI) is a $10^\circ \times 10^\circ$ square centered at the TeV gamma-ray centroid of HESS J1857+026 \citep[R.A. = 284.296$^\circ$, Dec. = 2.667$^\circ$;][]{2018A&A...612A...1H}.
To reduce the contamination from the Earth Limb, events with zenith angle greater than $90^\circ$ are eliminated.
The standard analysis software of {\it Fermitools}
\footnote{http://fermi.gsfc.nasa.gov/ssc/data/analysis/software/} is used with the instrumental response function (IRF) of P8R3{\_}SOURCE{\_}V3.
The models we used to describe the Galactic and isotropic diffuse emissions include {\tt gll\_iem\_v07.fits}, {\tt iso\_P8R3\_SOURCE\_V3\_v1.txt} and 
{\tt iso\_P8R3\_SOURCE\_V3\_PSF3\_v1} \footnote{http://fermi.gsfc.nasa.gov/ssc/data/access/lat/BackgroundModels.html}.
In addition, all sources in the incremental version of the fourth {\em Fermi}-LAT source catalog \citep[4FGL-DR3;][]{2020ApJS..247...33A,2022ApJS..260...53A} within a radius of 15$^\circ$ centered at HESS J1857+026 are included in the model.
And the binned maximum likelihood analysis method with {\tt gtlike} is applied.
During the fitting procedure, the spectral parameters of all sources located in the ROI are left free, together with the normalizations of the two diffuse backgrounds.

\subsection{Spatial Analysis} 
\subsubsection{Average Spatial Extension}
\label{sec:av_ex}
\begin{figure}[ht]
\centering
\includegraphics[angle=0,scale=0.35,height=0.25\textheight]{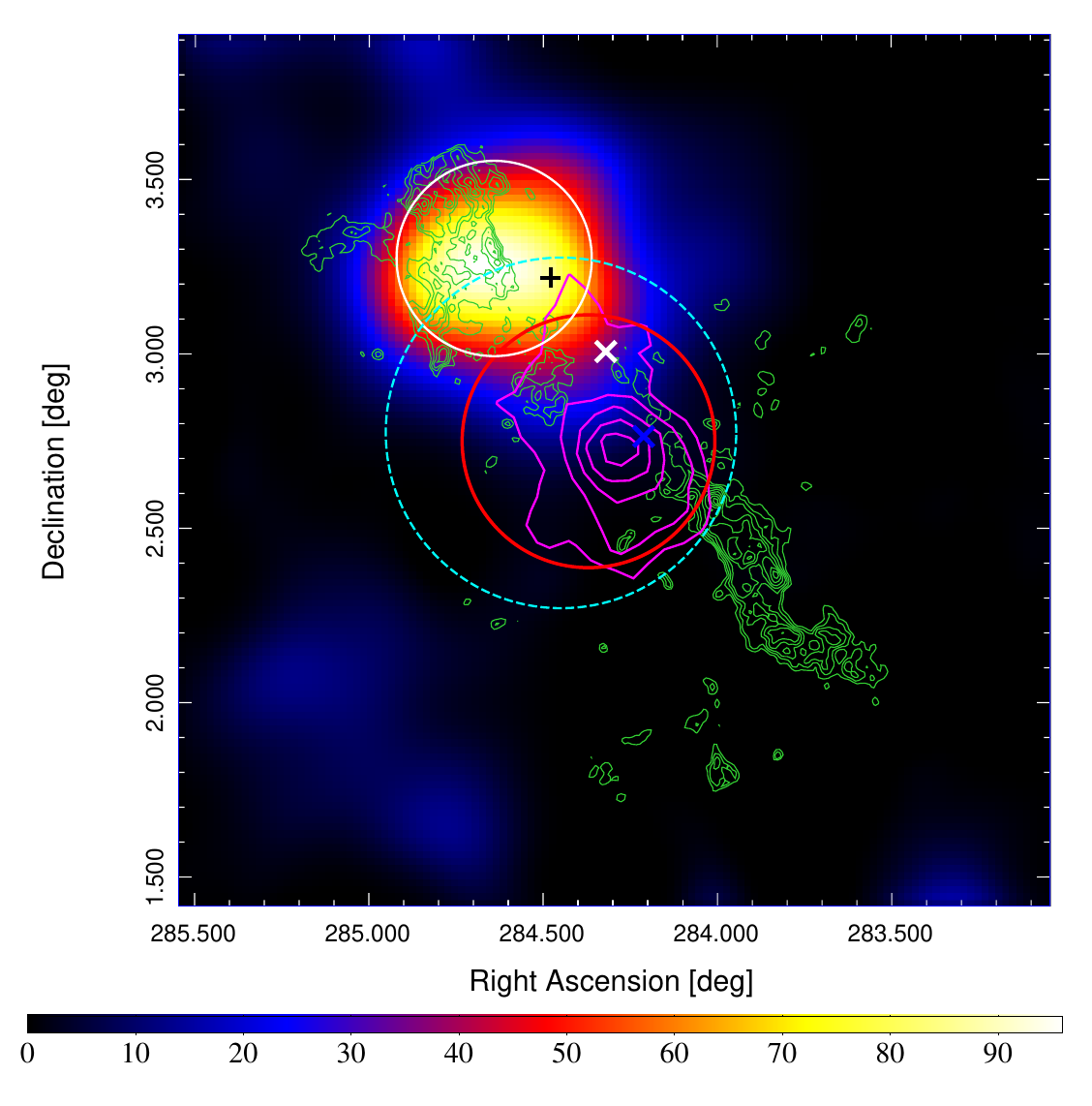}
\includegraphics[angle=0,scale=0.35,height=0.25\textheight]{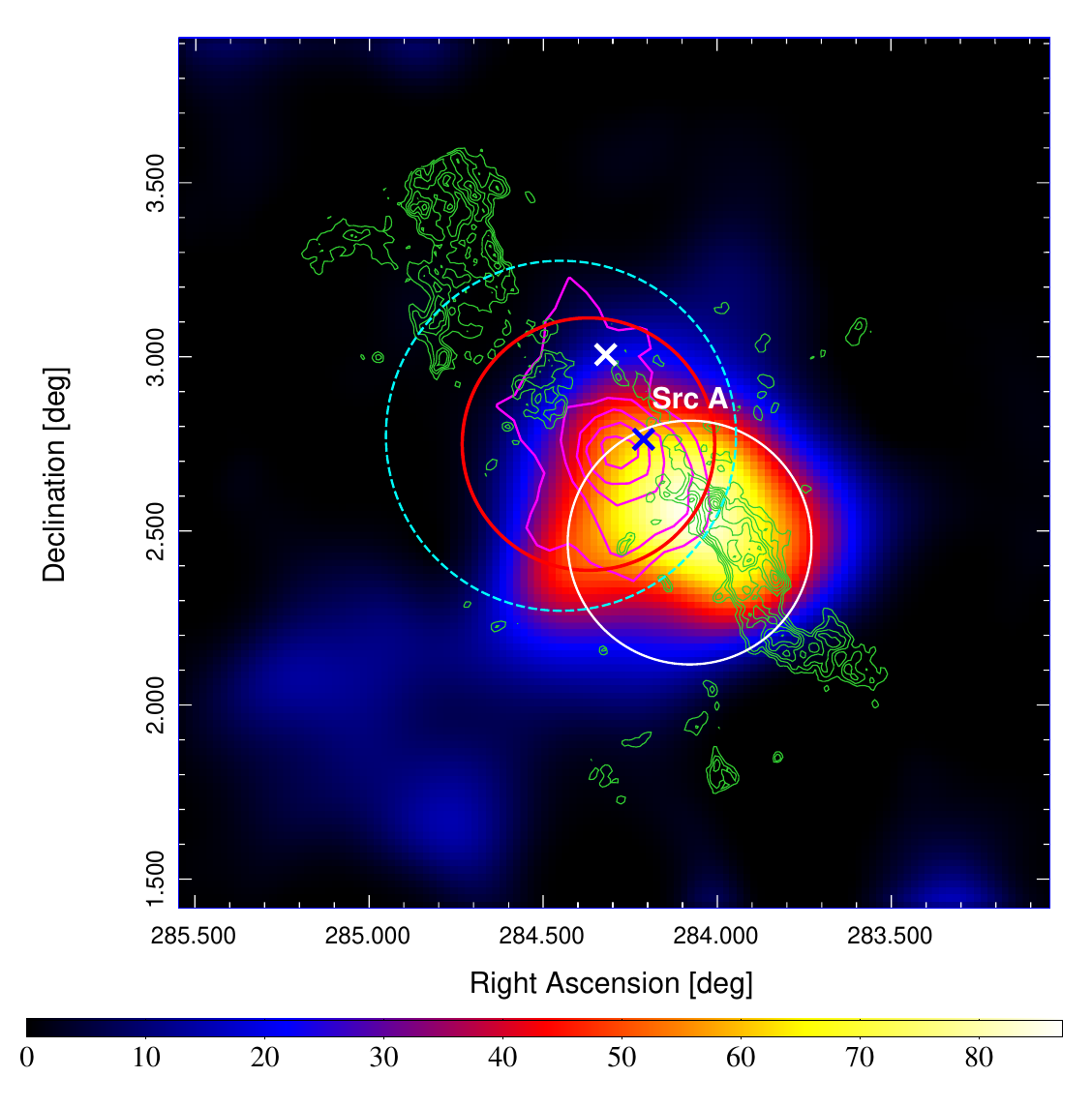}
\includegraphics[angle=0,scale=0.35,height=0.25\textheight]{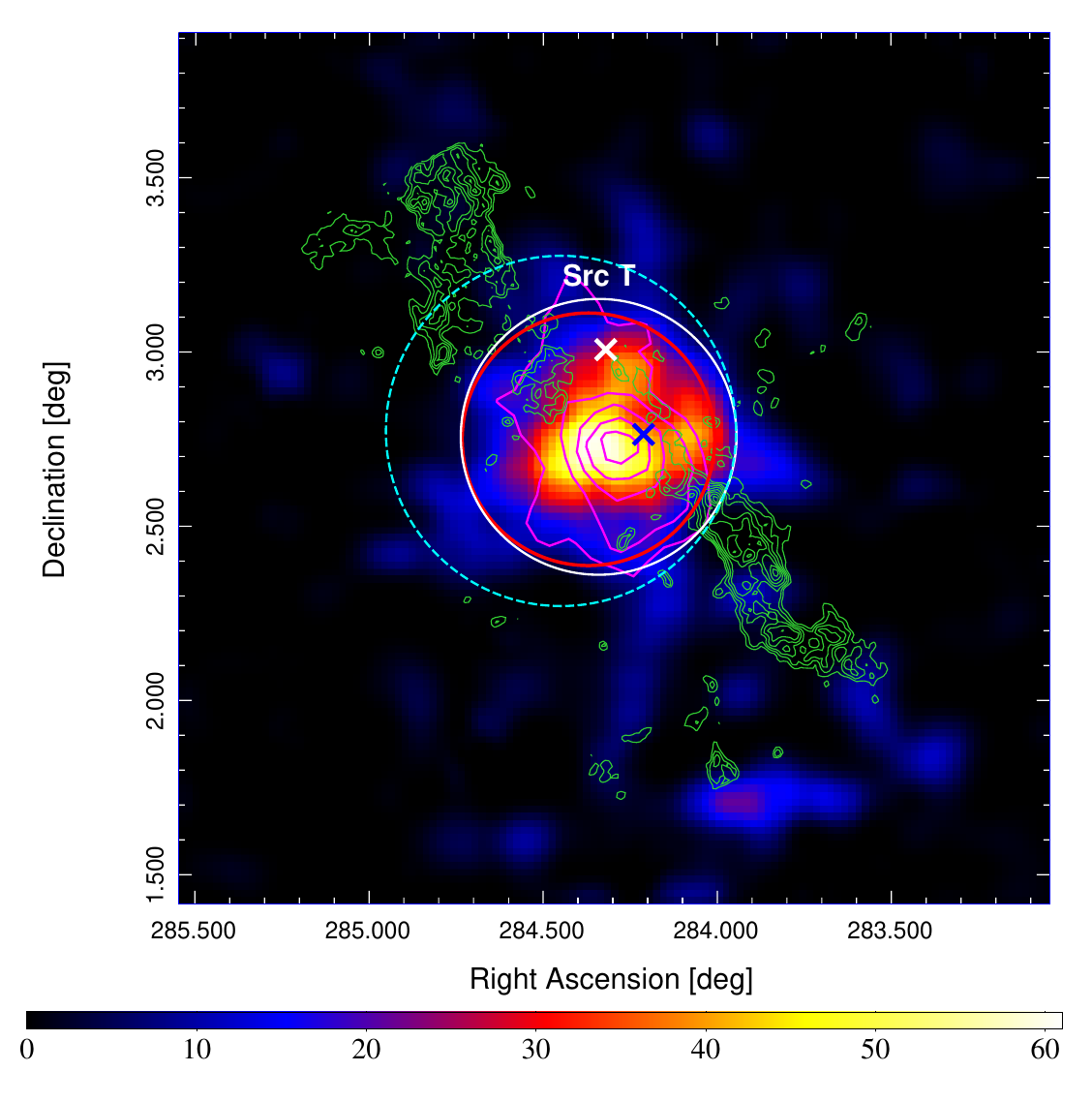}
\hfill
\caption{$2.5^\circ \times 2.5^\circ$ TS maps centered on HESS J1857+026. Each pixel is $0.02^\circ$ and the TS maps are smoothed with a Gaussian kernel of $\sigma=0.04^\circ$. 
The green contours indicate the $^{\rm 12}$CO($J$=1-0) emission	integrated between 78 and 90 $\rm{km}\ \rm{s}^{-1}$ by FUGIN \citep{2017PASJ...69...78U}.
The magenta contours are the H.E.S.S. observation of HESS J1857+026 \citep{2018A&A...612A...1H}. 
The cyan dashed circle is the $r_{\rm 68}$ of uniform disk of HESS J1857+026 given by FGES \citep{2017ApJ...843..139A}.
The red solid circle shows the $r_{\rm 68}$ of 2D Gaussian for 1LHAASO J1857+0245 detected by LHAASO-WCDA \citep{2023arXiv230517030C}.
And the blue and white crosses are the position of PSR J1856+0245 and PSR 1857+0300, respectively.
{\it Left:} TS map for the data below 3 GeV with the diffuse backgrounds and 4FGL-DR3 sources subtracted except for 4FGL J1857.9+0313e. 
The black plus shows the position of 4FGL J1857.9+0313c as a point source given by 4FGL-DR3, and the white circle sign the $r_{\rm 68}$ of best-fit uniform disk we analyzed here.
{\it Middle:} TS map for the data below 3 GeV with the diffuse backgrounds and 4FGL-DR3 sources subtracted except for Src A. 
The $r_{\rm 68}$ of the uniform disk for SrcA is described by the white circle.
{\it Right:} TS map with data above 10 GeV. 
The white circle marks the $r_{\rm 68}$ of the best-fit 2D Gaussian model for SrcT.}
\label{fig:tsmap_10gev}
\end{figure}

In the 4FGL-DR3 catalog, the gamma-ray emission of HESS J1857+026 is described by an uniform disk (named as 4FGL J1857.7+0246e) centered at R.A. = $284.449^\circ$, Dec. = $2.774^\circ$ with the 68\% containment radius of $r_{\rm 68}=0.50^\circ$, which is given by the analysis of {\it Fermi}-LAT extended Galactic sources \citep[FGES;][]{2017ApJ...843..139A}. 
Meanwhile, there is a point source named as 4FGL J1857.9+0313c located in the north of the disk, which has no identified counterpart \citep{2020ApJS..247...33A}.
And in the south of the disk, two point sources named as 4FGL J1857.6+0212 and 4FGL J1858.3+0209 are identified to be associated with SNR G35.6-0.4 and HESS J1858+020 \citep{2021A&A...646A.114C, 2022ApJ...931..128Z}.
To obtain the spatial template of HESS J1857+026, we performed the spatial extension analysis in the energy ranges of 1-3 GeV (low energy) and 10-500 GeV(high energy), respectively.
And in the low energy band, only ``PSF3'' type (evclass=128 \& evtype=32) data with better angular resolution are selected to reduce the contamination from nearby sources, 
while the data with ``SOURCE'' type (evclass=128 \& evtype=3) are used in the high energy range. 
After subtracting the background sources included in the model (except 4FGL J1857.9+0313c in left panel and 4FGL J1857.7+0246e in the middle and right panel), 
we created three $2.5^\circ \times 2.5^\circ$ TS maps centered at HESS J1857+026 with the different energy ranges, as shown in Figure \ref{fig:tsmap_10gev}. 


As shown in the left panel of Figure \ref{fig:tsmap_10gev}, the position of 4FGL J1857.9+0313c given by 4FGL-DR3 is not coincident with the the gamma-ray peak of this region.
In addition, there are discrepancies between the morphologies of the GeV emission around HESS J1857+026 and the spatial templates given by FGES both in the low and high energy bands, as shown in the middle and right panels of Figure \ref{fig:tsmap_10gev}.
Therefore, we refined the spatial templates of HESS J1857+026 of the low (hereafter referred to be ``Src A") and high (``Src T") energies, and the morphology of 4FGL J1857.9+0313c
was also re-analyzed.
The uniform disk, 2D Gaussian, as well as the H.E.S.S. image model are tested.
Meanwhile, a point source model with the best-fit coordinate calculated by {\tt gtfindsrc} is also applied for Src A and 4FGL J1857.9+0313c to 
test the spatial extension of it.
And the MAGIC image above 1 TeV is also adopted as the spatial template of Src T to explore whether or not the gamma-ray emission above 10 GeV is a superposition of two sources, which is similar to MAG1 and MAG2 \citep{2014A&A...571A..96M}.
The centroids and extensions of the 2D Gaussian and uniform disk are fitted by {\it Fermipy} \footnote{https://fermipy.readthedocs.io/en/latest/}, a {\tt PYTHON} package that automates analyses with the Fermi Science Tools \citep{2017ICRC...35..824W}.
And the results of the spatial analysis with the different energy ranges are listed in Table \ref{table:template}.
From Table \ref{table:template}, we can see that both the uniform disk and 2D Gaussian we analyzed can describe the GeV emission of Src A and 4FGL J1857.9+0313c (hereafter renamed as 4FGL J1857.9+0313e).
And for Src T, a 2D Gaussian model is the best-fit template instead of the MAGIC image, indicating that the GeV emission from Src T is in favor of one-source scenario instead of two separate gamma-ray sources.
In the following analyses, the uniform disk is adopted as the spatial template of Src A and 4FGL J1857.9+0313e, while Src T is described by the 2D Gaussian model.

As shown in the middle panel of Figure \ref{fig:tsmap_10gev}, the centroid of the gamma-ray emission from Src A is far from the position of PSR J1856+0245/PSR 1857+0300 and is on the edge of the TeV emission of HESS J1857+026, which suggests that there is no spatial coincidence between Src A and the pulsars or TeV emission.
For Src T, the extension of the best-fit 2D-Gaussian template is much smaller than the result given by FGES, which could be attributable to the improvement of the Galactic diffuse background or the newly detected gamma-ray sources.
Considering the comparable extension and the spatial coincidence between Src T and H.E.S.S. observation \citep{2018A&A...612A...1H},
we suggest that the GeV emission of Src T has the same origin of the TeV emission.
In addition, we found that the spectral index of Src A in the energy range of 1-3 GeV is $\Gamma=2.50 \pm 0.34$, while the spectral index of Src T for the data 
above 10 GeV is $\Gamma=1.84 \pm 0.10$. 
Considering the distinct GeV spectra and morphologies, the origins of Src A and Src T are probably different.
And both Src A and Src T are considered in the following analysis.
%

\begin{table*}[ht]
\centering
\caption {Spatial Properties for the GeV Emission in the Direction of HESS J1857+026}
\begin{tabular}{ccccccccc}
\hline \hline
Energy Range & Spatial Template & Sources & R.A.  & Dec.  & $r_{\rm 68} \tablenotemark{$\star$}$& $\Delta$log L \tablenotemark{$\dagger$} & d.o.f \\ 
             &    &    &(deg) & (deg) &  (deg)  &    & \\
\hline
1-3 GeV &  Point+Disk (4FGL)    &  4FGL J1857.9+0313c & $284.478$  & $3.219$ &  $-$   & {\multirow{2}{*}{0}}  & {\multirow{2}{*}{10}} \\
        &                       &  Src A              & $284.449$  & $2.774$ & $0.50$  &      &       \\
\cline{2-8}
        &   Two Point           &  4FGL J1857.9+0313c & $284.665$  & $3.260$ & $-$    & {\multirow{2}{*}{6}}  & {\multirow{2}{*}{9}}    \\
        &                       &  Src A              & $284.038$  & $2.589$ & $-$    &       &        \\
\cline{2-8}
        &  Point+Disk           &  4FGL J1857.9+0313c & $284.665$  & $3.260$ & $-$    & {\multirow{2}{*}{26}}  & {\multirow{2}{*}{10}}     \\
        &                       &  Src A              & $284.169 \pm 0.036$ & $2.504 \pm 0.047$ & $0.40 \pm 0.03$& &\\
\cline{2-8}
        &  Point+Gaussian       &  4FGL J1857.9+0313c & $284.665$  & $3.260$ & $-$    & {\multirow{2}{*}{25}}   & {\multirow{2}{*}{10}}    \\
        &                       &  Src A              & $284.178 \pm 0.045$ & $2.503 \pm 0.064$ & $0.51 \pm 0.07$& &\\
\cline{2-8}
        &  Two Disks            &  4FGL J1857.9+0313c & $284.641 \pm 0.035$  & $3.274 \pm 0.039$ & $0.28 \pm 0.04$& {\multirow{2}{*}{34}}& {\multirow{2}{*}{11}} \\
        &                       &  Src A              & $284.079 \pm 0.043$  & $2.467 \pm 0.041$ & $0.35 \pm 0.03$&     &        \\
\cline{2-8}
        &  Disk+Gaussian        &  4FGL J1857.9+0313c & $284.641 \pm 0.035$  & $3.274 \pm 0.039$ & $0.28 \pm 0.04$& {\multirow{2}{*}{32}}& {\multirow{2}{*}{11}} \\
        &                       &  Src A              & $284.028 \pm 0.046$  & $2.498 \pm 0.041$ & $0.31 \pm 0.05$&     &        \\
\cline{2-8}
        &  Gaussian+Disk        &  4FGL J1857.9+0313c & $284.641 \pm 0.039$  & $3.273 \pm 0.038$ & $0.29 \pm 0.06$& {\multirow{2}{*}{34}}& {\multirow{2}{*}{11}} \\
        &                       &  Src A              & $284.079 \pm 0.043$  & $2.467 \pm 0.041$ & $0.35 \pm 0.03$&     &        \\

\hline
10-500 GeV & Disk (4FGL)        & Src T     &   $284.449$           & $2.774$           & $0.50$          & $0$ & $5$ \\ 
           & Disk               & Src T     &   $284.361\pm 0.030$  & $2.797 \pm 0.029$ & $0.36 \pm 0.02$ & $15$ & $5$  \\
           & Gaussian           & Src T     &   $284.341 \pm 0.031$ & $2.757 \pm 0.035$ & $0.40 \pm 0.04$ &  $20$ & $5$  \\
           & H.E.S.S. Image     & Src T     &    $-$                & $-$               & $-$             & $15$  & $2$  \\ 
           & MAGIC Image        & Src T     &    $-$                & $-$               &  $-$            & $14$  & $2$\\
\hline
\hline
\end{tabular}
\tablenotetext{\star}{$r_{\rm 68}$ is the 68\% containment radius, where $r_{\rm 68}=1.51 \sigma$ for 2D Gaussian model and $r_{\rm 68}=0.82 \sigma$ for uniform disk model \citep{2012ApJ...756....5L}.}
\tablenotetext{\dagger}{Calculated with respect to the spatial model used in 4FGL-DR3.}
\label{table:template}
\end{table*}

\subsubsection{Energy-Dependent Extension Analysis of Src T}
\label{sec:ed_ex}
\begin{figure*}[!htb]
\centering
\includegraphics[height=0.25\textheight]{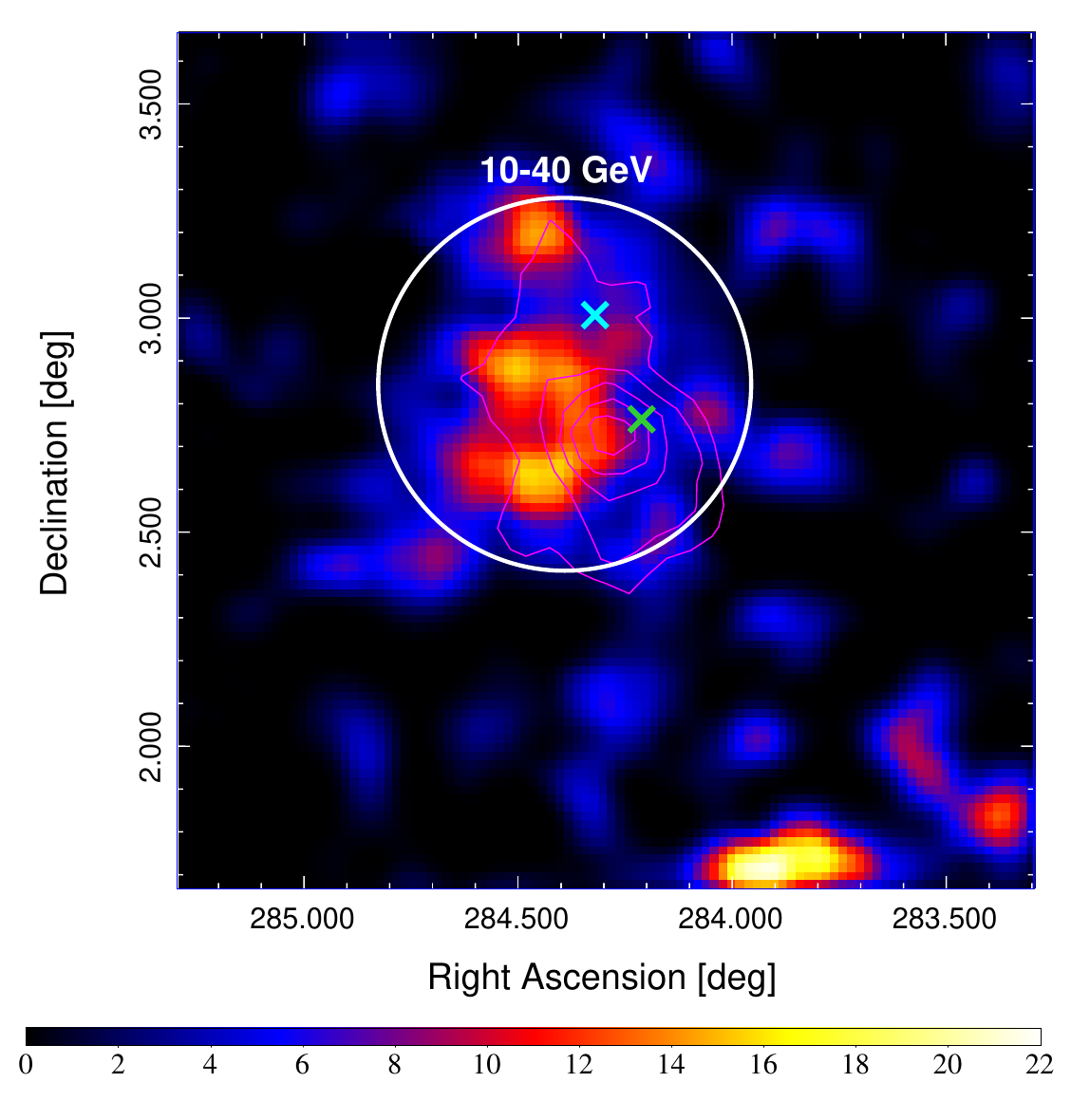}
\includegraphics[height=0.25\textheight]{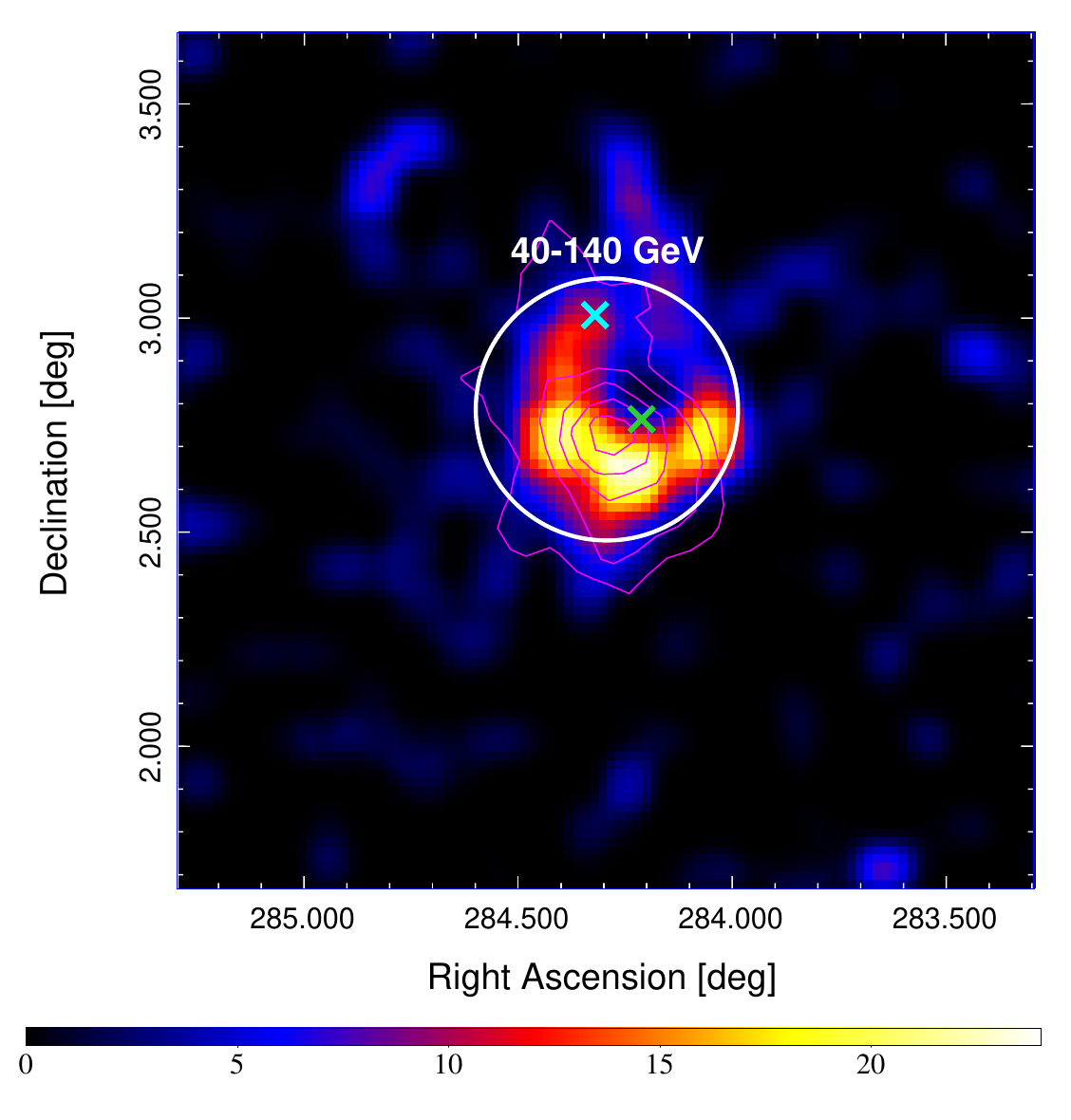}
\includegraphics[height=0.25\textheight]{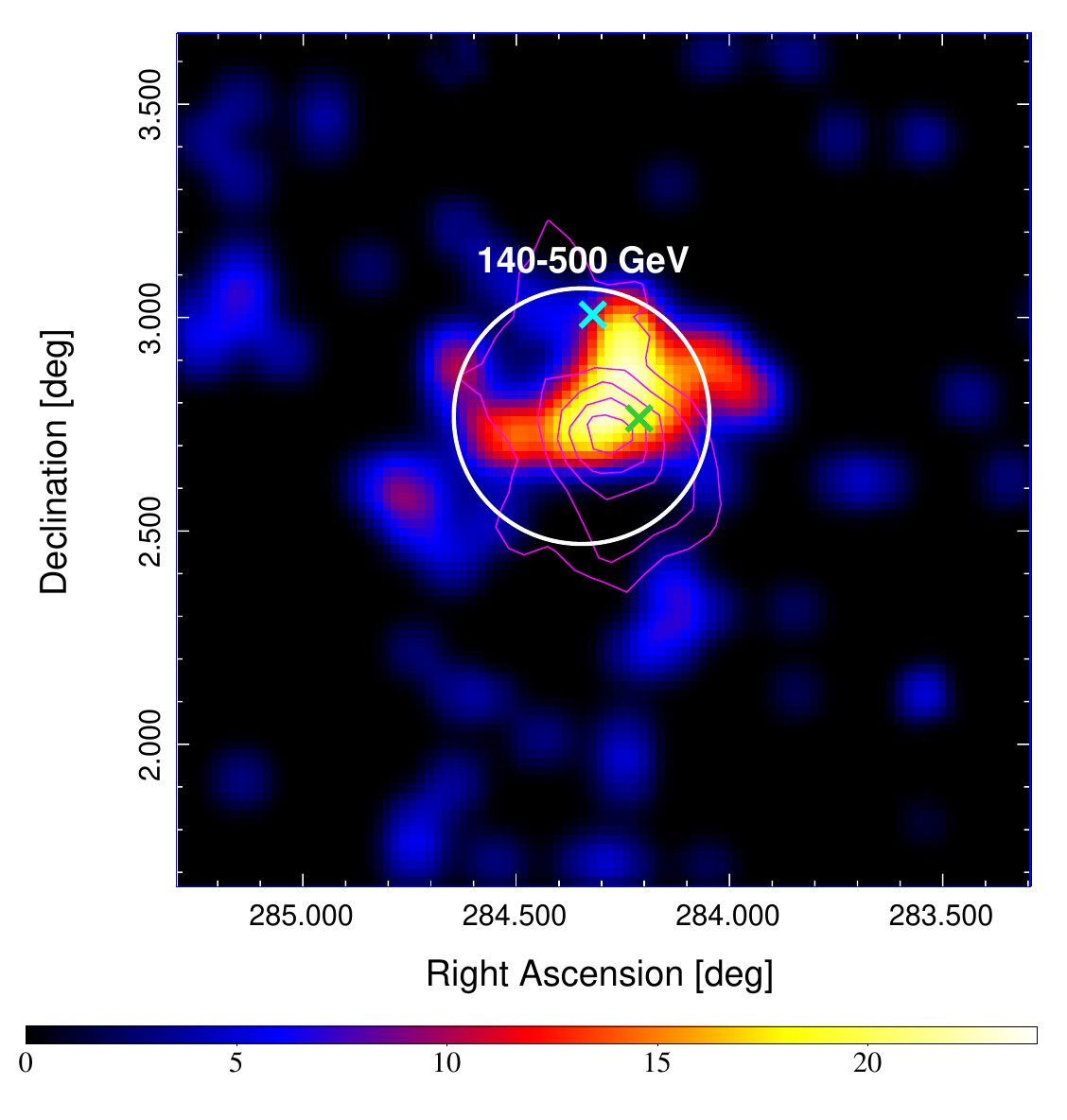}
\hfill
\caption{$2^\circ \times 2^\circ$ TS maps centered on HESS J1857+026 in the energy ranges of 10-40 GeV (left), 40-140 GeV (middle) and 140-500 GeV (right) . 
The white circles indicate the $r_{68}$ of 2D Gaussian template for the different energy bands.
The magenta contours are H.E.S.S. observation of HESS J1857+026 \citep{2018A&A...612A...1H}.
And the position of PSR J1856+0245 and PSR 1857+0300 are shown as the green and the cyan crosses, respectively.}
\vspace{3em}
\label{fig:e_dependent_tsmap}
\end{figure*}

To further explore the energy-dependent behavior of Src T, we performed the extension analyses in the energy ranges of 10-40 GeV, 40-140 GeV and 140-500 GeV, respectively.
For Src T, the 2D-Gaussian template is adopted, while the centroid and extension in each energy band are refitted with {\it Fermipy}.
The results with the different energy ranges are listed in Table \ref{table:e_dependent_ext}, and the corresponding TS maps are presented in Figure \ref{fig:e_dependent_tsmap}. 

The energy-dependent analysis show that the extension of Src T in the energy range of 10-40 GeV is larger than that of higher energy bands, with $r_{\rm 68}$ varying from $0.44^\circ$ below 40 GeV to $\sim 0.30^\circ$ above 140 GeV.
Similar phenomenons are also observed in the typical PWN HESS J1825-137 at TeV and GeV energies \citep{2006A&A...460..365A,2019A&A...621A.116H,2020A&A...640A..76P}. 
In addition, it should be noted that the gamma-ray emission regions in the energy range of 40-140 GeV and 140 GeV-500 GeV seem to be different, as shown in the middle and right panels of Figure \ref{fig:e_dependent_tsmap}. However, such phenomena could be attributed to the limited statistics of gamma-ray photons, and more observational data will be helpful to explore the energy-dependent behavior of Src T.


\begin{table}[!htb]
\centering
\caption {Extension Measurements of Src T in the Different Energy Ranges}
\begin{tabular}{cccccc}
\hline \hline
Energy Range & R.A.  & Dec.  & $r_{68}$   &TS  \\ 
                 & (deg) & (deg) &  (deg)  &   \\ 
\hline
10-40 GeV     &   $284.391 \pm 0.056$ & $2.846 \pm 0.070$ &  $0.44 \pm 0.07$ &$53$ \\ 
40-140 GeV      &   $284.293\pm 0.049$ & $2.787 \pm 0.053$& $0.31 \pm 0.05$ &$52$ \\ 
140-500 GeV  & $284.347 \pm 0.062$ &  $2.769 \pm 0.059$   & $0.30 \pm 0.06$ & $40$ \\ 
\hline
\hline
\end{tabular}
\label{table:e_dependent_ext}
\end{table}

\subsection{Spectral Analysis}
The spectral analysis was performed in the energy range of 500 MeV - 500 GeV.
And similar to the spatial analysis, the events of ``PSF3'' type with a better angular resolution were selected for the data below 3 GeV.
During the analysis process, the summed likelihood analysis method was adopted, and both Src A and Src T were included in the model with their spatial templates given in Sect. \ref{sec:av_ex}.
The spectra of Src A, Src T and 4FGL J1857.9+0313e are adopted to be the power law models.
And we also tested the GeV spectral curvature for each source by adopting the log-parabola spectrum, 
while no significant improvement was obtained compared with the power law models.
The global fitting gives a hard spectrum for Src T with the photon index of $\Gamma = 1.74 \pm 0.07$, 
which is comparable to the result in \citet{2012A&A...544A...3R}. And the integral photon flux is estimated to be 
$(5.04 \pm 1.08) \times 10^{-9}$ ph cm$^{-2}$ s$^{-1}$.
While the spectrum of Src A is very soft, and the photon index is fitted to be $\Gamma = 2.73 \pm 0.10$, 
with the integral photon flux of $(1.98 \pm 0.19) \times 10^{-8}$ ph cm$^{-2}$ s$^{-1}$.
And the best-fit photon index of 4FGL J1857.9+0313e is $\Gamma= 2.55 \pm 0.07$ with the integral photon flux of
$(1.41 \pm 0.13) \times 10^{-8}$ ph cm$^{-2}$ s$^{-1}$.

\begin{figure*}[!htb]
\centering
\includegraphics[width=0.48\textwidth]{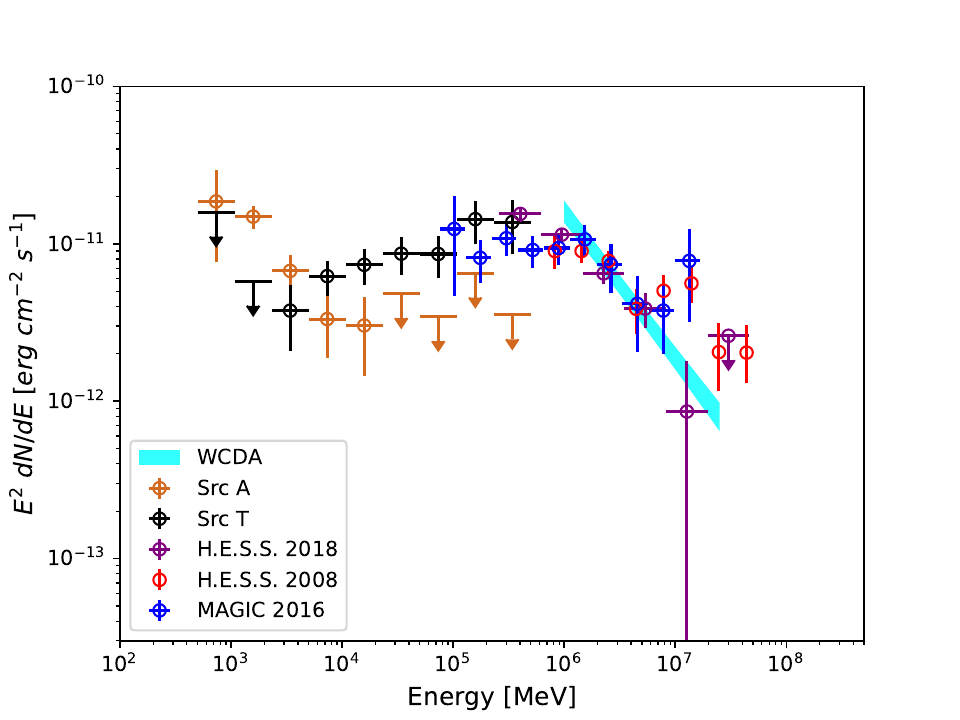}
\includegraphics[width=0.48\textwidth]{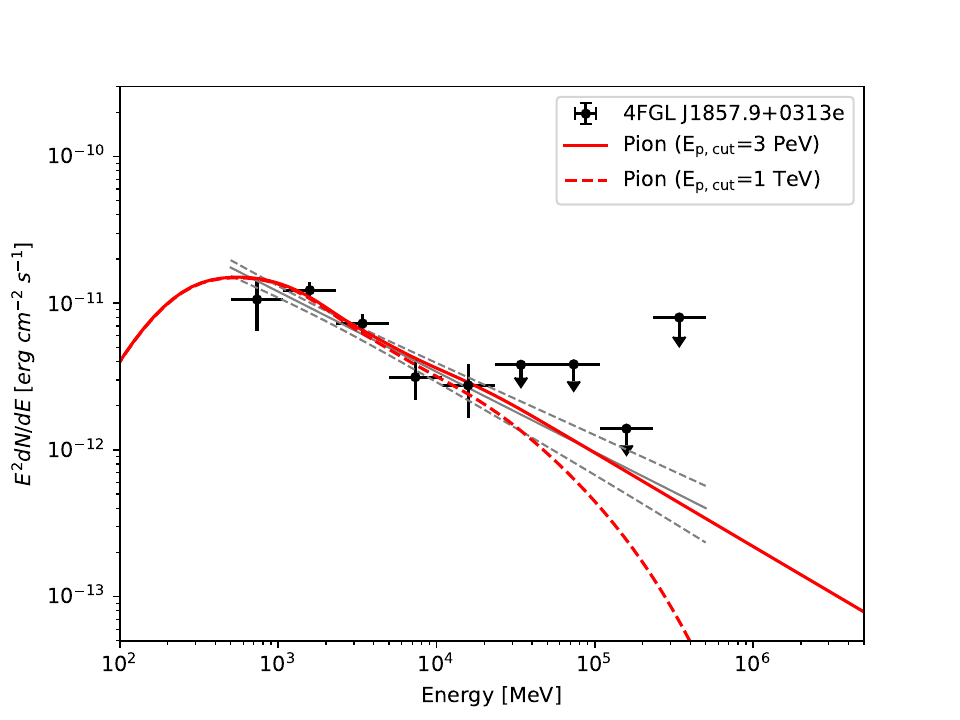}
\hfill
\caption{ {\it Left:} The GeV SEDs of SrcA and SrcT are marked by the brown and black circles, respectively.
The red and purple circles indicate the H.E.S.S. observations by \citet{2008A&A...477..353A} and \citet{2018A&A...612A...1H}, respectively. 
And the blue circles show the MAGIC observation in \citet{2014A&A...571A..96M}. And the cyan butterfly indicates the global power law spectrum 
of 1LHAASO J1857+0245 detected by LHAASO-WCDA \citep{2023arXiv230517030C}. {\it Right:} The GeV SED of 4FGL J1857.9+0313e. 
The gray solid and dashed lines show the best-fit power law spectrum and its 1$\sigma$ statistic error. And the hadronic models with $E_{\rm p,cut}$ = 3 PeV and $E_{\rm p,cut}$ = 1 TeV for 4FGL J1857.9+0313e are shown as the red solid and dashed lines, respectively.}
\label{fig:sed_srcT}
\end{figure*}

To study the spectral energy distributions (SEDs) of these three sources, the data are divided into nine logarithmically space energy bins.
And the summed likelihood analysis is repeated in each energy bin, with only the normalizations of sources located in ROI and the diffuse backgrounds in the model left free, while the spectral parameters are fixed to be the global-fitting values. 
For the energy bin with TS value lower than 4.0,
an upper limit with 95\% confidence level is calculated.
And the SEDs are shown in Figure \ref{fig:sed_srcT}. 
The GeV spectrum of Src T could connect with the TeV SED of HESS J1857+026 smoothly, which suggests the same physical origin. 

\section{Discussion}

\begin{figure*}[!htb]
\centering
\includegraphics[width=4in]{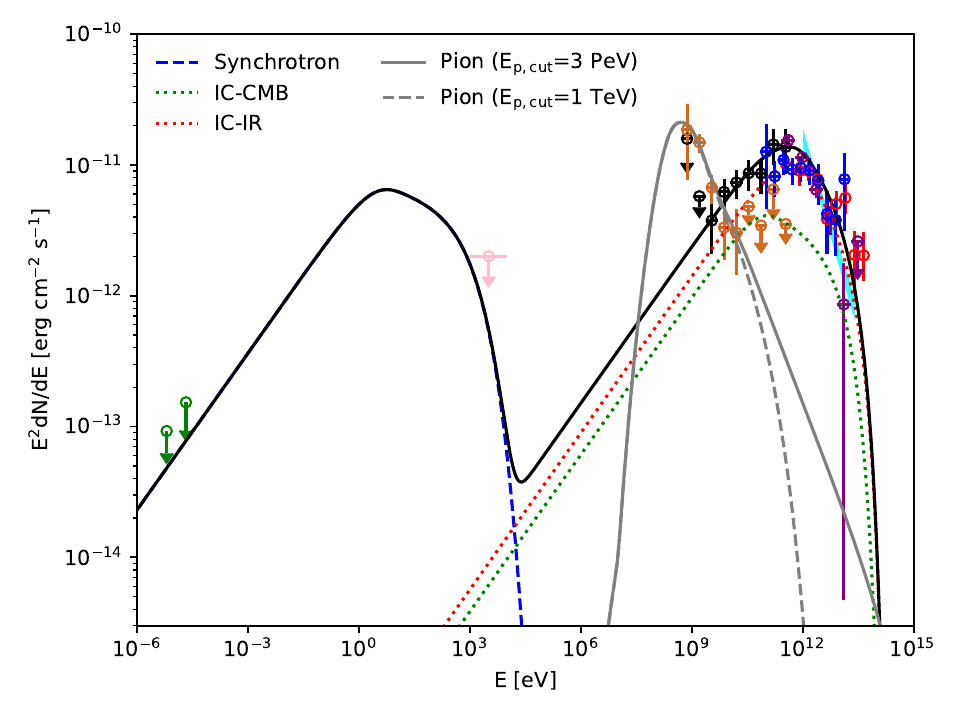}
\hfill
\caption{The broadband SED of HESS J1857+026 with the leptonic model and SrcA with hadronic model.
The radio and X-ray flux upper limits are shown as the green and pink dots \citep{2012A&A...544A...3R,2021A&A...652A.142P}.
The blue dashed line shows the synchrotron component for HESS J1857+026.
The red and green dotted lines represent IC scattering of different seed photons. 
And the black solid curve is the sum of different leptonic radiation components for HESS J1857+026.
The gray solid and dashed lines indicate the hadronic models with $E_{\rm p,cut}$ = 3 PeV and $E_{\rm p,cut}$ = 1 TeV for SrcA.}
\label{fig:multi_fit}
\end{figure*}

The spatial and spectral data analyses above reveal that the diffuse GeV gamma-ray emission around HESS J1857+026 could be distinguished into two separate extended components: one of them, namely Src A, has no spatial and spectral coincidence with HESS J1857+026. And another one, namely Src T, shows both spatial and spectral consistency with the TeV measurement of HESS J1857+026, which supports Src T as the GeV counterpart of HESS J1857+026. 

Although MAGIC observation revealed two sources in the region of HESS J1857+026, MAG1 and MAG2, with the events above 1 TeV \citep{2014A&A...571A..96M}, there is no evidence show that Src T is composed by two gamma-ray sources limited by the event statistics and the PSF of {\em Fermi}-LAT. Therefore, we suggest that the GeV emission of HESS J1857+026 originates from a single gamma-ray source.
The right panel of Figure \ref{fig:tsmap_10gev} shows that the centroid of GeV gamma-ray emission is consistent with PSR J1856+0245 associated with MAG1, not PSR J1857+0300 in the direction of MAG2, which supports the same origin for HESS J1857+026 and MAG1.
And for the possible origin of the gamma-ray emission from HESS J1857+026, \citet{2014A&A...571A..96M} suggested a PWN powered by PSR J1856+0245 with $\dot E$ = $4.6 \times 10^{36}$ erg s$^{-1}$, and the spin-down luminosity of PSR J1857+0300 with $\dot E$ $\sim 2.5 \times 10^{32} \ {\rm erg}\ {\rm s^{-1}}$ is too low to power a gamma-ray PWN \citep{2013ApJS..208...17A,2013ApJ...773...77A}.
While \citet{2021A&A...652A.142P} revealed the existence of a superbubble with the analysis of atomic gas in this region, and suggested the superbubble origin for the TeV emission from HESS J1857+026.
However, the hard GeV gamma-ray spectrum of HESS J1857+026 makes it to be different from the typical superbubble with an index of $\sim$ 2.2, e.g. Cygnus Cocoon \citep{2011Sci...334.1103A,2019NatAs...3..561A}, but is similar to the typical PWNe, e.g. MSH 15-52 \citep{2010ApJ...714..927A}, 
HESS J1825-137 \citep{2011ApJ...738...42G}. 
In addition, the GeV gamma-ray emission of HESS J1857+026 also shows the energy-dependent morphology, with the emission radius varying from $0.44^\circ$ below 40 GeV to $\sim 0.30^\circ$ above 140 GeV. 
And the centroid of the GeV emission moves towards PSR J1856+0245 with increasing energies.
Such characteristics are also detected in the PWN HESS J1825-137 and HESS J1303-631 \citep{2006A&A...460..365A,2019A&A...621A.116H,2012A&A...548A..46H,2020A&A...640A..76P}.
All these evidences above support that the gamma-ray emission of HESS J1857+026 could originate from a PWN associated with PSR J1856+0245.

For PWNe, the emission from radio to X-rays are normally produced by the synchrotron emission, whereas the gamma-ray emission are explained by the 
IC scattering process (leptonic process). 
Here, a simple one-zone leptonic model is applied for HESS J1857+026.
The IC background photon fields include the cosmic microwave background (CMB) and infrared (IR) photon from dust with the temperature of T $\sim$ 30 K and density of 1 eV cm$^{-3}$ \citep{2012A&A...544A...3R}.
And the distance of HESS J1857+026 is adopted to be 6.3 kpc, derived from the dispersion measurement of PSR J1856+0245 \citep{2017ApJ...835...29Y}.
Considering the absence of radio detection for the PWN, we used the sensitivity of VLA in \citet{2021A&A...652A.142P} to calculate the upper limits of radio flux by assuming that the spatial extension of the radio PWN is same to the that of the gamma-ray emission with a radius of $0.4^\circ$. And the radio upper limits are estimated to be 9.2$\times$10$^{\rm -14}$ erg cm$^{\rm -2}$ s$^{\rm -1}$ at 1.5 GHz and 1.5$\times$10$^{\rm -13}$ erg cm$^{\rm -2}$ s$^{\rm -1}$ at 5 GHz, respectively.
The electron spectrum is assumed to be a broken power law with an exponential cutoff in the form of
$\frac{dN_e}{dE} \propto  \frac{(E/E_{br})^{-\alpha_1}}{1+(E/E_{br})^{\alpha_2-\alpha_1}}\rm{exp} \left(-\frac{E}{E_{e,cut}} \right)$ \citep{2011ApJ...738...42G,2018ApJ...867...55X},
where $E_{\rm br}$, $E_{\rm e,cut}$, $\alpha_1$ and $\alpha_2$ are the break energy, cutoff energy, and indices of electron spectrum, respectively.
The model fitting is performed using the {\it naima} package \citep{2015ICRC...34..922Z}. 

As shown in Figure \ref{fig:multi_fit}, the gamma-ray spectrum of HESS J1857+026 can be reproduced with $\alpha_1 \sim 2.2$, $\alpha_2 \sim 3.2$, a break 
energy with $E_{\rm br} \sim 3.5$ TeV, and a cutoff energy with $E_{\rm e,cut} \sim 70$ TeV, respectively.
The total energy of electrons above 1 GeV, $W_e$, is calculated to be $\sim 1.1 \times 10^{49} (d/6.3\ {\rm kpc})^2$ erg.
The cooling timescale at break energy is estimated to be $t_{\rm cool} \approx 56 ({\rm E}/3.5\ {\rm TeV})^{-1} [(U_{\rm ph}+U_{\rm B})/1.66\ {\rm eV} \ {\rm cm}^{-3}]^{-1}$ kyr, here $U_{\rm ph}$ = 1.26 eV/${\rm cm}^3$ and $U_{\rm B}= B^2/8\pi$. 
And such value is about two times larger than the characteristic age of PSR J1856+0245, 
which suggests that the break structure could be an intrinsic characteristic of injected electronic spectrum, not be produced by the cooling effect \citep{2006ARA&A..44...17G}.
With the electron spectrum, the radio and X-ray upper limits constrain the magnetic field strength to be lower than $\sim$4$\mu$G, which is consistent with the typical values for gamma-ray PWNe \citep[e.g.][]{2011ApJ...738...42G,2013ApJ...774..110G}.
It should be noted that the X-ray upper limit was calculated with a radius of $0.1^\circ$ by \citet{2012A&A...544A...3R}. And with the same radius of $0.4^\circ$ for the calculation of radio upper limits, the X-ray flux would be much larger, which would have no affect to the SED fitting.

Along with the evolution of the PWN into the interstellar medium (ISM), the energetic electrons could escape and transport dominated by diffusion to form a detectable halo around the pulsar, which is
defined as a pulsar halo. Such halos were first detected around Geminga and PSR B0656+14 with the TeV gamma-ray emission \citet{2017ApJ...843...40A}.
And \citet{2019PhRvD.100l3015D} also claimed to detect the corresponding GeV gamma-ray emission around Geminga. 
Based on the definition of an electron halo in \cite{2020A&A...636A.113G}, 
namely that of the overdensity of relativistic electrons around pulsar compared with ISM,
we estimate the electronic energy density of HESS J1857+026, $\varepsilon_{\rm e}$, and compared it with the typical value of ISM, $\varepsilon_{\rm ISM}$ = 0.1 eV cm$^{-3}$.
The gamma-ray emission region of HESS J1857+026 is assumed to be a sphere with a radius of $\sim 0.40^\circ$, which 
corresponds to a physical radius of $\sim$ 44 pc for the distance of 6.3 kpc. Using the total energy of electrons we derived, 
the electronic energy density of $\varepsilon_{\rm e}$ $\sim 0.60 \ {\rm eV} \ {\rm cm}^{-3}$ is obtained, which is much larger than that of ISM.
Therefore, we suggest that the relativistic electrons are still contained in a region energetically and dynamically dominated by the pulsar.

\begin{figure*}[!htb]
	\centering
	\includegraphics[width=0.46\textwidth]{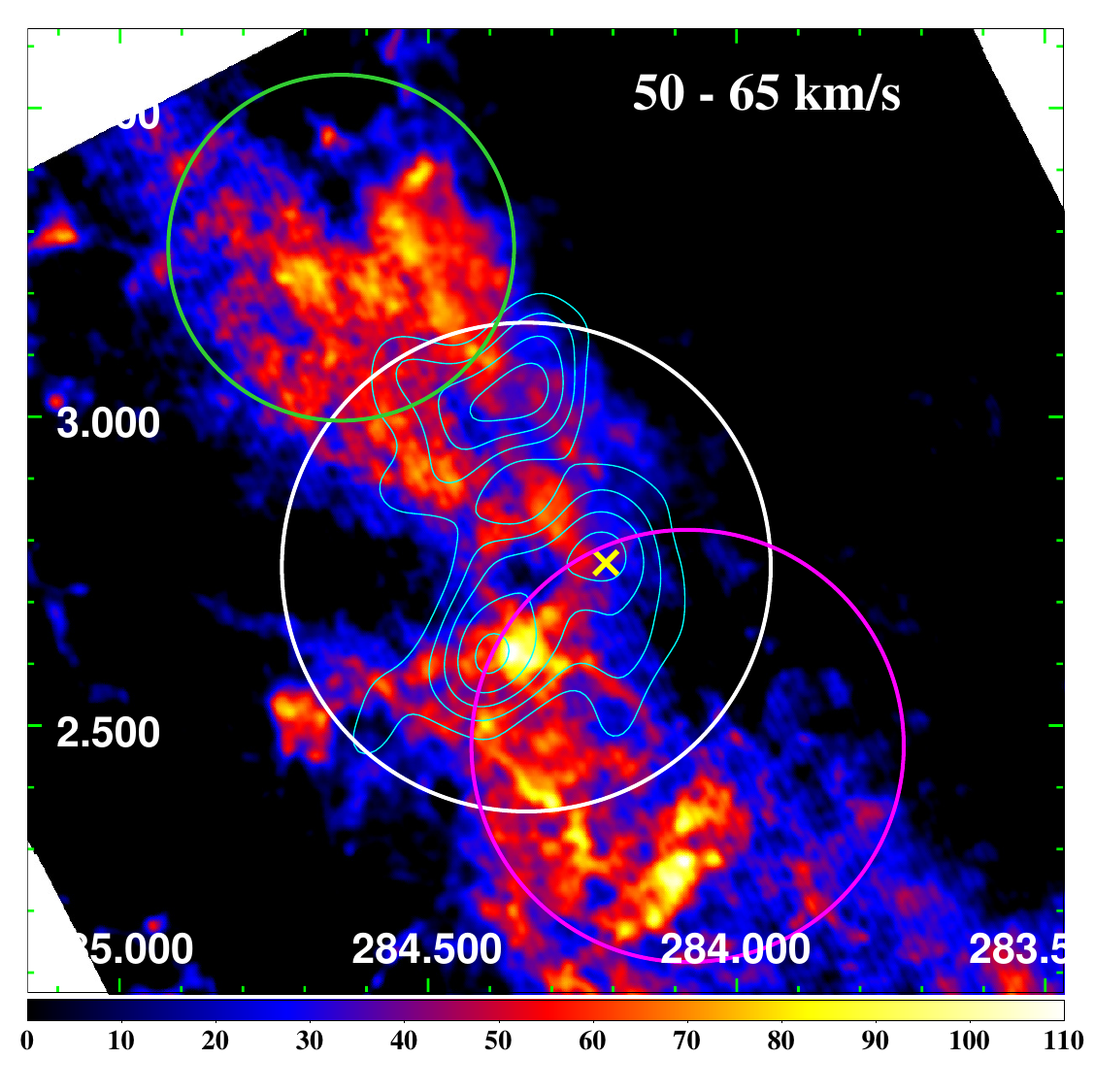}
	\includegraphics[width=0.46\textwidth]{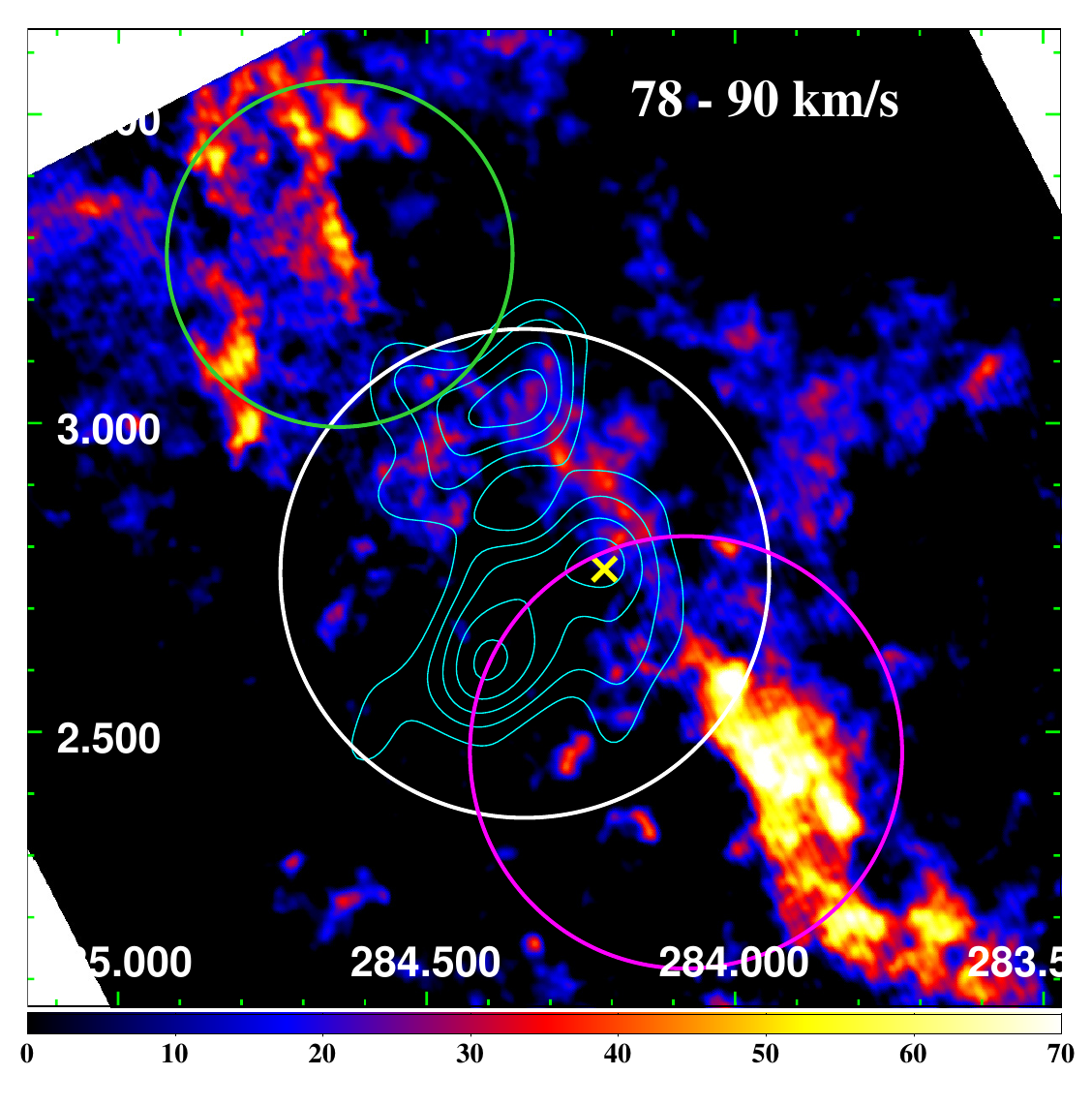}
	\hfill
	\caption{$^{12}$CO($J$=1–0) intensity maps (in the unit of K km s$^{-1}$) in the velocity ranges of 50-65 km s$^{-1}$ (left) and 78-90 km s$^{-1}$ (right). The green, white and magenta circles show the GeV gamma-ray extensions of SrcA, HESS J1857+026, and 4FGL J1857.9+0313e, respectively. The position of PSR J1856+0245 is marked by the yellow cross. And the cyan contours indicate the MAGIC gamma-ray flux map above 1 TeV \citep{2014A&A...571A..96M}.}
	\label{fig:co}
\end{figure*}

For the other two extended gamma-ray sources, SrcA and 4FGL J1857.9+0313e, either of them is not spatially consistent with the identified counterpart in other wavelengths.
By analyzing the $^{\rm 12}$CO($J$=1-0) line data performed by the FOREST Unbiased Galactic plane Imaging survey with the Nobeyama 45 m telescope \citep[FUGIN;][]{2017PASJ...69...78U}, two molecular clumps in the southwest (named as clumpA) and northeast (named as clumpB) of HESS J1857+026 are founded in both the velocity ranges of 50 - 65 $\rm{km}\ \rm{s}^{-1}$ and 78 - 90 $\rm{km}\ \rm{s}^{-1}$, as shown in Figure \ref{fig:co}, which are spatially coincidence with the GeV gamma-ray emission from SrcA and 4FGL J1857.9+0313e, respectively.
The molecular clouds in the range of 50 - 65 $\rm{km}\ \rm{s}^{-1}$ have been revealed in \citet{2014A&A...571A..96M} with corresponding kinetic distance of $\sim$3.7 kpc.
And \cite{2021A&A...652A.142P} studies the molecular clouds in the range of 79 - 90 $\rm{km}\ \rm{s}^{-1}$ and derived a kinetic distance of $\sim$5.5 kpc, which is compatible, within the errors, with the DM distance of PSR J1856+0245 ($\sim$6.3 kpc).
By adopting the value of the conversion factor of $X_\mathrm{CO} = 2\times10^{20}$ cm$^{-2}$ (K km s$^{-1})^{-1}$ \citep{2013ARA&A..51..207B}, we estimate the total mass contents of clumpA and clumpB with the different distances, respectively.
And the total masses of clumpA and clumpB are estimated in the region of $0.35\degr$ and $0.28\degr$ sky integration radii, corresponding to the 68\% containment radii of the extended gamma-ray emission of SrcA and 4FGL J1857.9+0313e. respectively.
For the velocity range of 50 - 65 $\rm{km}\ \rm{s}^{-1}$ with distance of 3.7 kpc, the  total mass contents of clumpA and clumpB are calculated to be $\sim$ $2.1\times 10^{5}d_{3.7}^{2} M_{\odot}$ and $\sim$ $1.7\times 10^{5}d_{3.7}^{2} M_{\odot}$, corresponding to the average gas number densities of $\rm{n_{\rm gas,A}}$ = 175 cm$^{-3}$ and $\rm{n_{\rm gas,B}}$ = 280 cm$^{-3}$ by assuming a spherical geometry of the gas distribution.
For the velocity range of 79 - 90 $\rm{km}\ \rm{s}^{-1}$ with the compatible distance of PSR J1856+0245 with 6.3 kpc, the total mass content of clumpA and clumpB are calculated to be $\sim$ $1.5\times 10^{5}d_{6.3}^{2} M_{\odot}$ and $\sim$ $1.6\times 10^{5}d_{6.3}^{2} M_{\odot}$. And the corresponding average gas number densities are about $\rm{n_{\rm gas,A}}$ = 27 cm$^{-3}$ and $\rm{n_{\rm gas,B}}$ = 50 cm$^{-3}$, respectively.
The spatial coincidence between the extended gamma-ray emission and the molecular gas suggest the hadronic origin for SrcA and 4FGL J1857.9+0313e.

In the hadronic scenario, the proton spectrum is assumed to be a single power law with an exponential cutoff in the form of 
$\frac{dN_p}{dE} \propto E^{-\gamma} \rm{exp} \left(-\frac{E}{E_{p,cut}} \right)$.
And the cutoff energy of protons cannot be well constrained and was first adopted to be the energy of the cosmic ray knee with $E_{\rm p,cut}$ = 3 PeV.
The hadronic model for Src A is shown as the gray solid line in Figure \ref{fig:multi_fit}, and the corresponding proton spectrum should be much soft with $\gamma \sim 2.8$. 
The total energy of protons above 1 GeV is estimated to be $W_p \sim 1.6 \times 10^{50}(n_{\rm gas,A}/27\ {\rm cm}^{-3})^{-2}(d/6.3\ {\rm kpc})^2$ erg or $W_p \sim 1.6 \times 10^{48}(n_{\rm gas,A}/175\ {\rm cm}^{-3})^{-2}(d/3.7\ {\rm kpc})^2$ erg.
For 4FGL J1857.9+0313e, the spectral index of protons with $\gamma \sim 2.7$ and total energy of $W_p \sim 5.8 \times 10^{49}(n_{\rm gas,B}/50\ {\rm cm}^{-3})^{-2}(d/6.3\ {\rm kpc})^2$ erg or $W_p \sim 6.4 \times 10^{47}(n_{\rm gas,B}/280\ {\rm cm}^{-3})^{-2}(d/3.7\ {\rm kpc})^2$ are needed to explain the GeV gamma-ray emission, which is shown as the red solid line in the right panel of Figure \ref{fig:sed_srcT}.
In addition, considering the fact that the cut-off energy of protons in the middle-aged SNRs is usually much lower than PeV \citep{2013ApJ...777..148G,2023A&A...679A..22S}, we decreased the cutoff energies of protons to estimate the different hadronic models. And the allowed minimum value of cutoff energy is about 1 TeV for SrcA and 4FGL J1857.9+0313e, with the total energies of protons of  $\sim 1.5 \times 10^{50}(n_{\rm gas,A}/27\ {\rm cm}^{-3})^{-2}(d/6.3\ {\rm kpc})^2$ erg and $\sim 5.7 \times 10^{49}(n_{\rm gas,B}/50\ {\rm cm}^{-3})^{-2}(d/6.3\ {\rm kpc})^2$ erg, respectively.

For the molecular cloud in 50 -65 $\rm{km}\ \rm{s}^{-1}$ with the distance of 3.7 kpc, there is no candidates as the origin of high energy protons. While for the molecular clouds in 79 - 90 $\rm{km}\ \rm{s}^{-1}$, the compatible distance with PSR J1856+0245 suggests that the hypothetical SNR associated with PSR J1856+0245 could provide enough power assuming that $\sim$10\% of the supernova kinetic energy of $\sim$10$^{\rm 51}$ erg is transferred to the energy of particles.
Moreover, the soft GeV gamma-ray spectra of SrcA and 4FGL J1857.9+0313e are also similar to that of the old SNRs interacting with molecular clouds
\citep[e.g. IC 443 \& W44][]{2010Sci...327.1103A,2013Sci...339..807A}.
The absence of the detection of associated SNR in this region suggests that it may have already dissipated into the ambient gas.
And further observations, especially in the radio and X-ray bands, will be crucial to reveal the physical origin of the gamma-ray emission in this region.

\section{Conclusion}
Using more than 13 years of {\it Fermi}-LAT observations, we studied the GeV gamma-ray emission in the direction of HESS J1857+026, and found that the GeV emission around HESS J1857+026 is composed of two extended gamma-ray sources: Src A and Src T. 
Src T is spatially coincident with HESS J1857+026 and its hard GeV gamma-ray spectrum could connect with the TeV SED of HESS J1857+026 smoothly, 
indicating that SrcT could be the GeV counterpart of HESS J1857+026. 
In addition, we performed the energy-dependent analyses of the GeV gamma-ray emission from HESS J1857+026, and its extension decreases 
towards higher energies. 
The energy-dependent morphology and the hard GeV gamma-ray spectrum of HESS J1857+026 make it favor the PWN origin.
And one-zone leptonic model with a broken power law electronic spectrum can well describe the multi-wavelength data of HESS J1857+026. 
SrcA and another extended gamma-ray source, 4FGL J1857.9+0313e with soft GeV gamma-ray spectra, have no identified counterparts in other wavelength.
However, two molecular clumps in the northeast and southwest of HESS J1857+026 are spatially coincidence with the GeV gamma-ray emission from SrcA and 4FGL J1857.9+0313e, which suggest the hadronic process for their gamma-ray emission.
With the single power law model for protons, the GeV spectra of SrcA and 4FGL J1857.9+0313e could also be explained with the soft proton spectra.
And these high energy protons could be produced by the hypothetical SNR associated with PSR J1856+0245, which may have already dissipated into the ambient gas.

HESS J1857+026 is one of the peculiar gamma-ray sources, which shows the energy-dependent morphology.
More detailed observations by LHAASO \citep{2019arXiv190502773C} and CTA \citep{2019scta.book.....C} would be helpful to explore the particle transport mechanisms, 
and the future radio and X-ray observations are crucial to investigate the origin of the gamma-ray emission in this region.


\section*{Acknowledgments}
We acknowledge the use of the {\it Fermi}-LAT data provided by the Fermi Science Support Center.
This work is supported by the National Natural Science Foundation of China under the grants 12103040, 12147208 and U1931204, and the Natural Science Foundation for Young Scholars of Sichuan Province, China (No. 2022NSFSC1808).
%




\bibliography{ms}
\bibliographystyle{aasjournal.bst}



\end{document}